\documentclass{article}
\usepackage[english]{babel}
\usepackage[utf8]{inputenc}
\usepackage{johd}
\usepackage{wrapfig}
\usepackage{float} 

\title{Appropriateness of the McNamara and Buland's (2004) methodology for computing frequency-dependent seismic power}

\author{Sebin John$^{a}$$^{*}$ and Michael E. West$^{a}$ \\
        \small $^{a}$University of Alaska Fairbanks \\\\
        \small $^{*}$Corresponding author: Sebin John; \tt{sjohn19@alaska.edu} \\
}
\date{}

\begin{document}
\raggedright
\maketitle

\begin{abstract}
The methodology developed by McNamara and Buland (2004) for computing Power Spectral Densities (PSDs) has gained popularity due to its low computational cost and reduction of spectral variance. This methodology is widely used in seismic noise studies and station performance evaluations and is implemented in tools like ISPAQ, MUSTANG, and PQLX. However, concerns have been raised about its appropriateness in certain contexts, particularly when high-resolution spectral detail is required. This study evaluates McNamara and Buland’s methodology by comparing it with Welch’s method across three Alaskan stations with differing microseism conditions. When calculating seismic power across a band of frequencies--for example, the 5-10s secondary microseism--we find that both methodologies produce time series with nearly identical trends, albeit with slight differences in absolute power values. Our results demonstrate that McNamara and Buland’s methodology is fully appropriate for certain applications, specifically ones that rely on averaged seismic energy over a frequency band as opposed to a single discrete frequency.
\end{abstract}

\section*{\large Background}

\paragraph*{} The advent of digital recording systems in the 1980s and increased processing capabilities led to the widespread use of Power Spectral Densities (PSDs) in seismic noise source studies (\citealp{Withers1996}; \citealp{Young1996}; \citealp{Aster2008}; \citealp{Tsai2011}), station and network performance analysis (\citealp{Given1990}; \citealp{Peterson1993}; \citealp{Koymans2021}) and evaluating seismic instrument performance (\citealp{Holcomb1989}; \citealp{Sleeman2006}).\begin{wrapfigure}{l}{0.55\textwidth}
\includegraphics[width=0.5\textwidth]{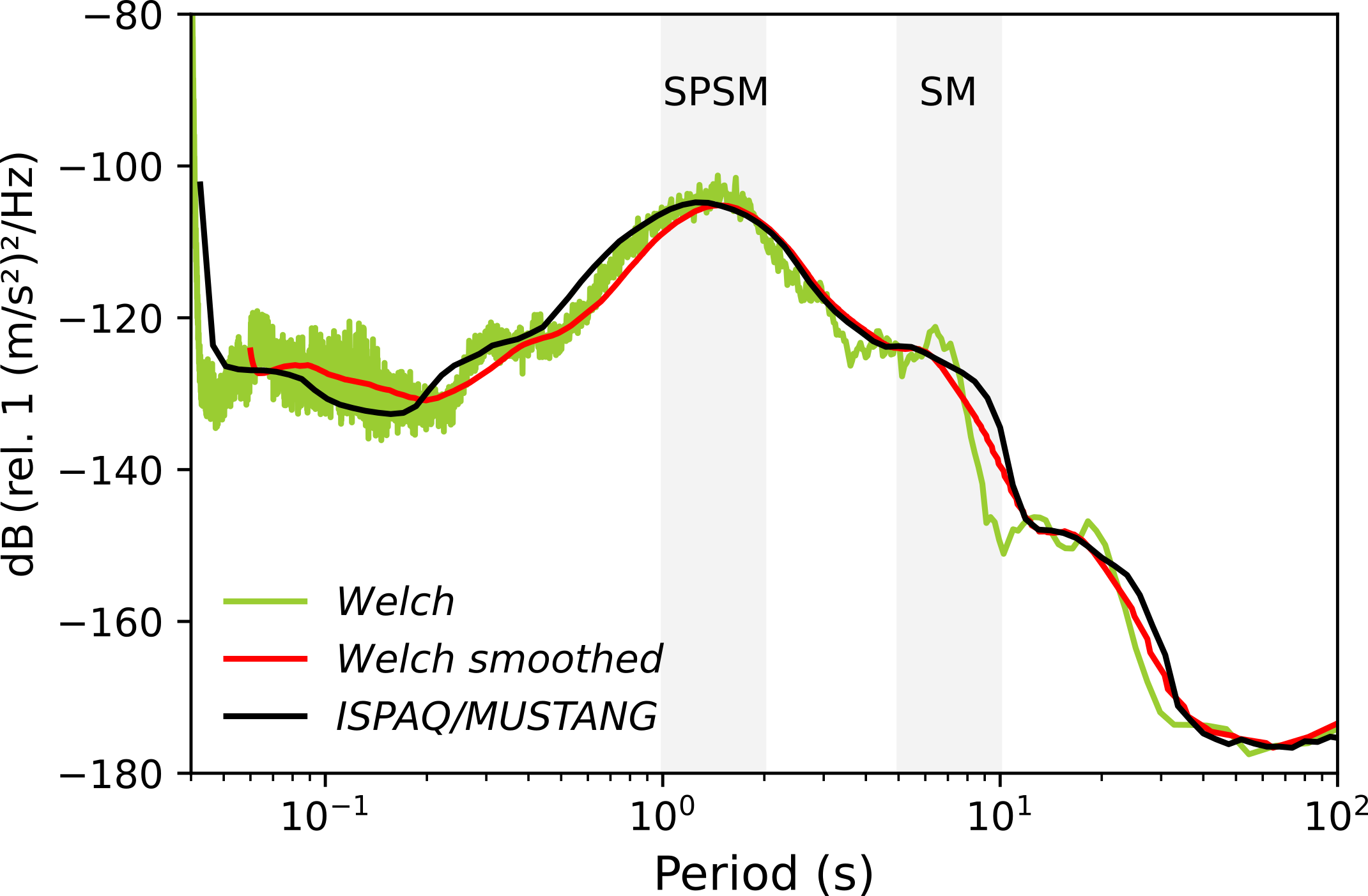}
\caption{\label{fig1}Spectra obtained using the Welch's method and ISPAQ. Redline shows one octave smoothed spectra computed using Welch's method}
\end{wrapfigure} Computing PSDs for long time series is still computationally expensive and produces large data volumes. Most subsequent analyses distill these data by stacking or averaging in either time or frequency. McNamara and Buland's methodology for computing PSDs (\citealp{McNamara2004}) has proven popular because of its low computational cost and the reduction of spectral variance. Several widely-
used software packages for computing PSDs use this methodology, including Modular Utility for STAtistical kNowledge Gathering (MUSTANG) (\citealp{Casey2018}), IRIS System for Portable Assessment of Quality (ISPAQ) (\citealp{Casey2018}), Obspy PPSD and PASSCAL Quick Look eXtended ( PQLX) (\citealp{McNamara2004}). 

\paragraph*{}However, \citealp{Anthony2020} raise several cautions about this methodology. They illustrate how smoothing across frequency windows with a one-octave width may compromise analyses that require high-resolution spectral information. They demonstrate that in some cases, the smoothed spectra can fall outside the 95\% confidence interval of spectra derived from the conventional Welch's method.\begin{wrapfigure}{l}{0.5\textwidth}
\includegraphics[width=0.5\textwidth]{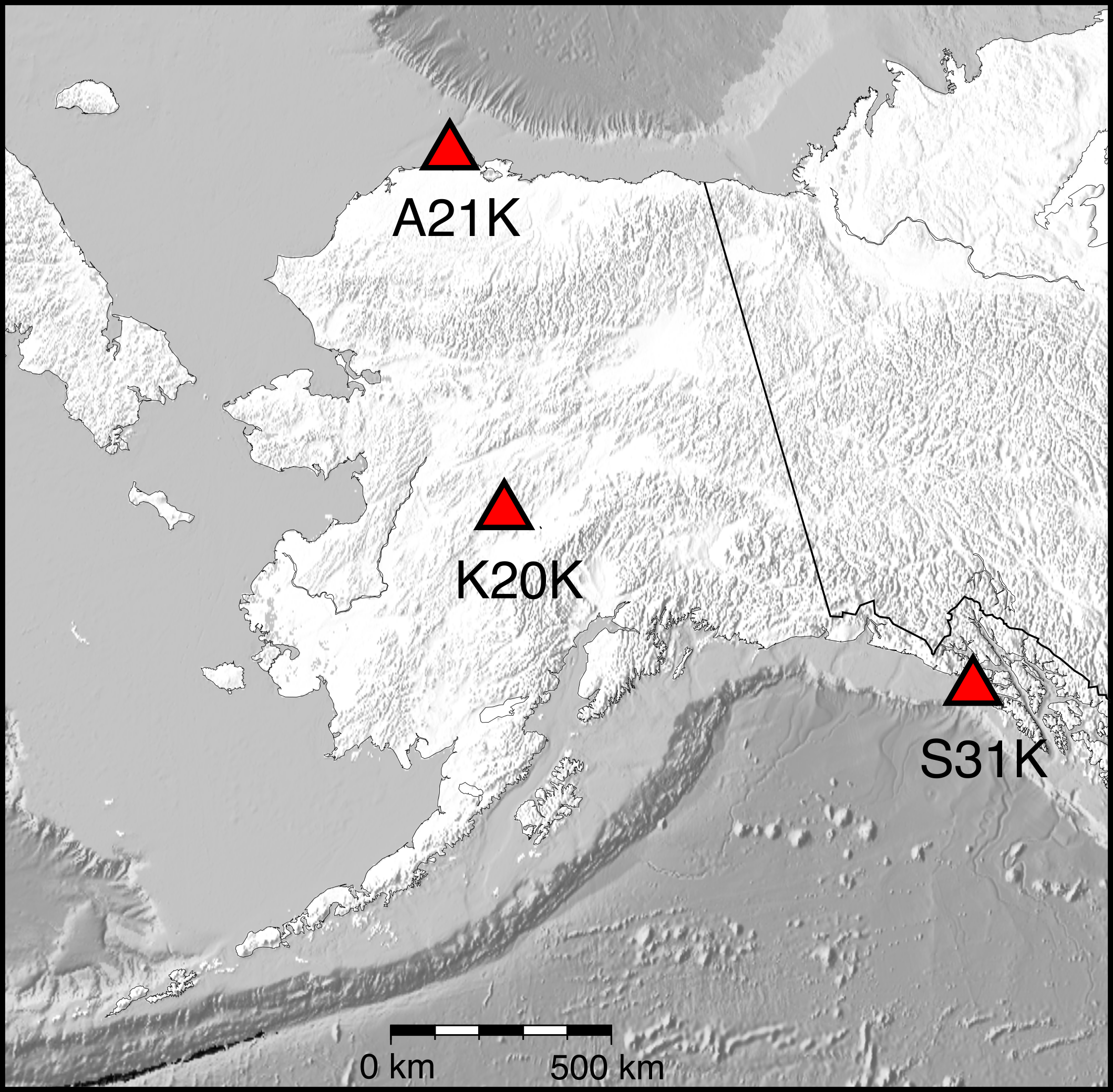}
\caption{\label{fig2}Location of different stations used for this analysis}
\end{wrapfigure} Earthquakes are also not removed in McNamara and Buland's methodology on the argument that earthquakes are low-probability occurrences in the background of ambient seismic noise. \citealp{Anthony2020} showed that teleseismic earthquakes can bias PSDs in the period range 10–50s. For these reasons, there is some concern about when these spectral products may be appropriate for research applications and when they may not be.

\paragraph*{} We demonstrate that McNamara and Buland's methodology is fully appropriate for certain applications, specifically those that rely on averages across spectral bands spanning over an octave or more. Fig.1 shows hourly spectra computed for station A21K (Fig.2) using the conventional segment-averaged Welch method (\citealp{Welch1967}) and ISPAQ, which uses McNamara and Buland's methodology. For this study, we focus on secondary microseism bands between 5-10s and 1-2s. The ~5-10s secondary microseism band is of particular interest because seismic power can vary significantly between closely spaced frequencies in this range. We hereafter refer to mean PSDs between 5-10 seconds as Secondary Microseismic (SM) power and between 1-2s as Short Period Secondary Microseismic (SPSM) power. 

\section*{\large Analysis }

\paragraph*{} To quantify the difference between PSDs obtained from Welch's method and that computed using McNamara and Buland's methodology, we select three stations located in regions of Alaska (Fig.2) with very different microseism conditions and calculate PSDs using both methods. We use the ISPAQ module to compute PSDs using McNamara and Buland's (2004) methodology. 

\paragraph*{} To compute PSDs without octave smoothing, we divide the data into hourly segments and apply Welch's (\citealp{Welch1967}) section-averaging method with sub segments of five and a half minutes with 50\% overlap. We remove the instrument response from the resulting spectra and convert them to decibels for one-to-one comparison with the output from ISPAQ.

\paragraph*{} For each hour, we compute an average power value in the SM and SPSM bands using both methodologies. The result are time series of power values derived from the two different PSD methodologies. A comparison of the SM time series from both methodologies for the year 2021 is shown in Fig.3. A closer look at the month of May is also shown in the right column of the figure. Fig.4 is the corresponding figure for the SPSM band. The times series from the two methodologies are highly similar, though not identical. The long-wavelength seasonal trends mirror one another. Variations on the scale of days to weeks, reflecting the influence of individual storms, are also highly similar. We do note very different power values for individual outlier hours. These outliers are, without exception, above the main trend and represent time segments compromised by earthquakes, anthropogenic noise, or instrumental issues. We see no consistent patterns in how these outliers are handled by the two approaches. More importantly however, we also note that these are specifically data points that should be excluded from most downstream analysis. Regardless of the methodology, these are outlier data that are not faithfully capturing the background continuous noise.

\paragraph*{} We can measure the similarity of time series from both methodologies through cross-correlation. Table 1 shows cross-correlation coefficients for each pair of year-long time series in three very different noise environments. Cross-correlation coefficients are close to 1 for both the SM and SPSM bands, demonstrating that the power time series obtained from both methodologies are extremely similar in shape. In other words, the relative amplitude between peaks and troughs in both signals is almost the same. This illustrates that the McNamara and Buland's methodology can be used reliably to track broad variations in seismic power through time. The reason that both approaches generate such similar time series, despite the variations illustrated in Fig.1, is because we are taking the mean over a band of frequencies, effectively smoothing the signal. In fact, our SM and SPSM bands are one octave in width (5-10s and 1-2s, respectively)--the same smoothing introduced by McNamara and Buland. 

\begin{table}[H]
	\label{table1}
	\centering
	\begin{tabular}{llll}
		\hline
		Band width    & A21K   & S31K & K20K \\
		\hline
		5-10s (SM)     & 0.98      & 0.99 & 0.99 \\
		1-2s (SPSM)     & 0.99    & 0.99 & 0.93 \\
		\hline
	\end{tabular}
  	\caption{Cross-correlation coefficients between time series obtained using Welch's and ISPAQ methodology}
\end{table}%

\paragraph*{} Though both methodologies produced PSD time series of similar shape, there is some consistent difference in absolute seismic power (Fig.3 \& Fig.4).  PSD time series obtained using ISPAQ show slightly higher amplitudes compared to Welch's method, irrespective of station location or time of year. To analyze this difference, we calculate the difference in mean seismic power between the two methodologies (Fig.5 \& Fig.6). The distribution of this difference is shown in the second column of Fig.5 and Fig.6. This distribution follows a Gaussian curve with a slight negative bias for both bands. The offset at the three stations we observe is on average 1.4-2.2 db for the SM band and 0.3-4.2 db for the SPSM band. While these offsets are real, they are far smaller than the $\sim40$ db variations we track annually. 

\section*{\large Conclusions}

\paragraph*{} We conclude that it is fully appropriate to use tools (and datasets pre-computed and readily available) using the McNamara and Buland's (2004) methodology for certain applications. Specifically, we find it a fine choice for tracking seismic power through time–a use case where it is almost always preferable to track trends in a well-chosen frequency band as opposed to at a single discrete frequency. We highlight a few caveats above. But so long as these are kept in mind, the cost/benefit of using readily available frequency-smoothed spectral power values may be advantageous for many research purposes. 

\begin{figure}[H]
\centering
\includegraphics[width=1\textwidth]{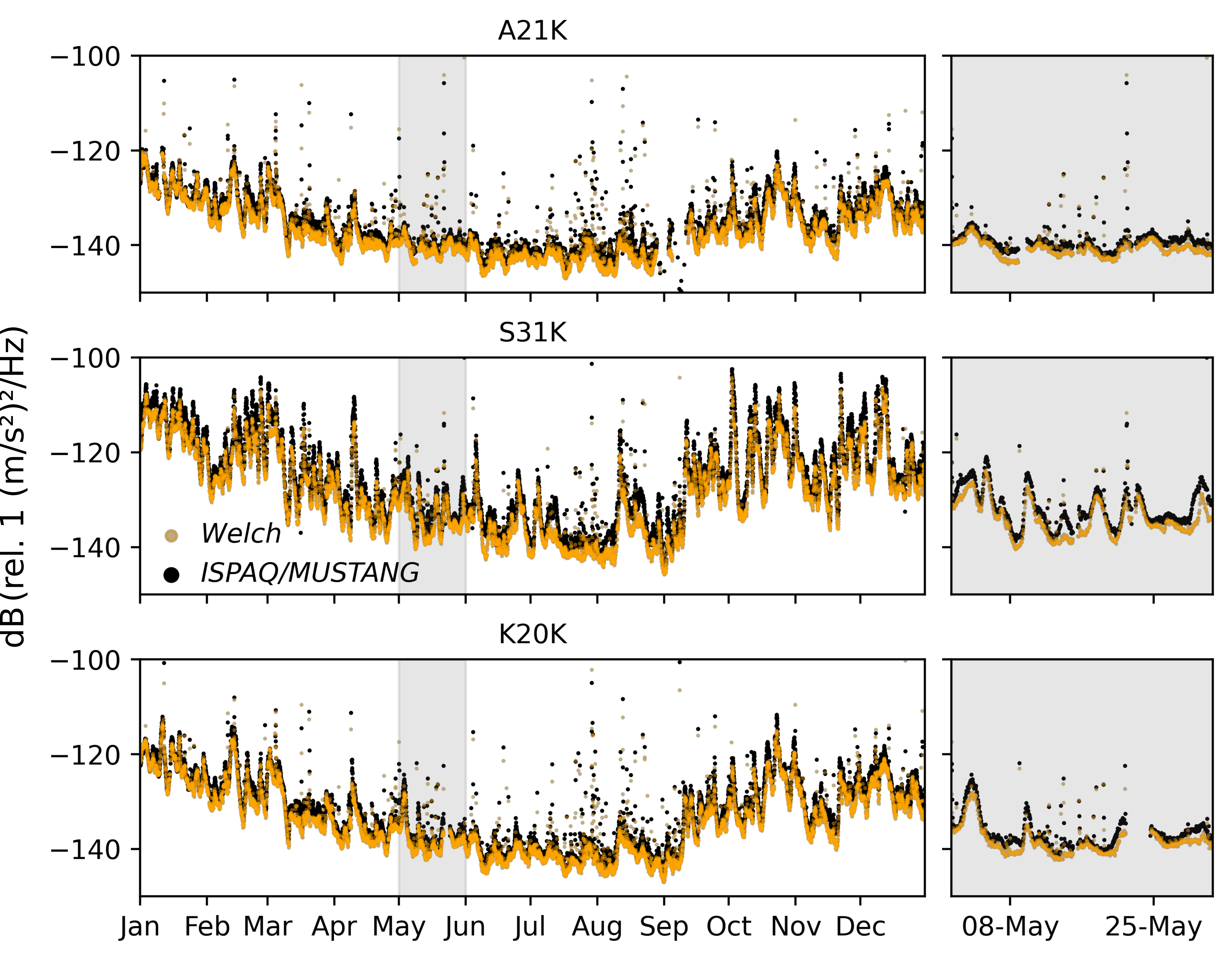}
\caption{\label{fig3}Mean PSDs in the SM (5-10s) band computed using ISPAQ/MUSTANG and Welch's method.}
\end{figure}

\begin{figure}[H]
\centering
\includegraphics[width=1\textwidth]{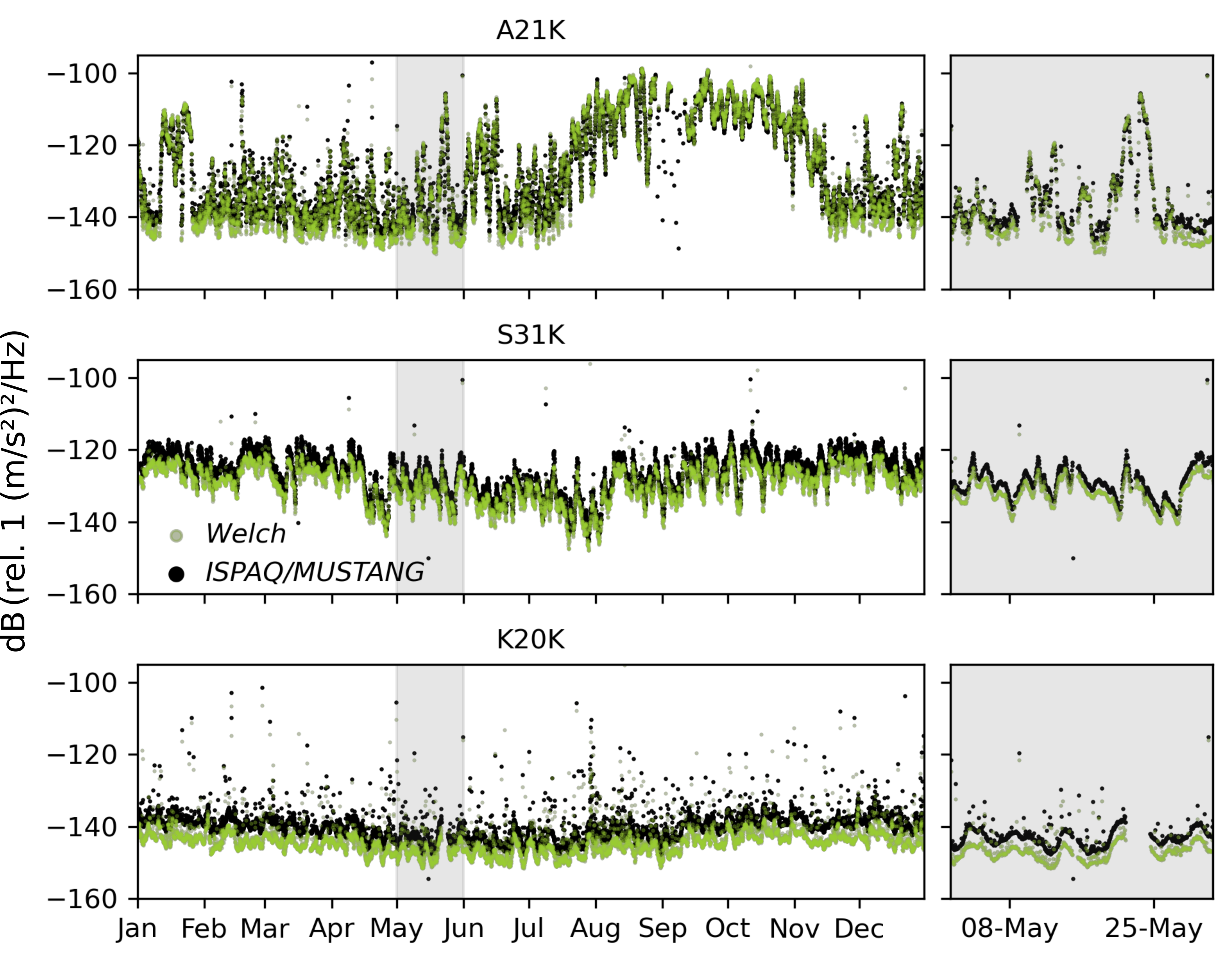}
\caption{\label{fig4}Mean PSDs in the SPSM (1-2s) band computed using ISPAQ/MUSTANG and Welch's method.}
\end{figure}

\begin{figure}[H]
\centering
\includegraphics[width=1\textwidth]{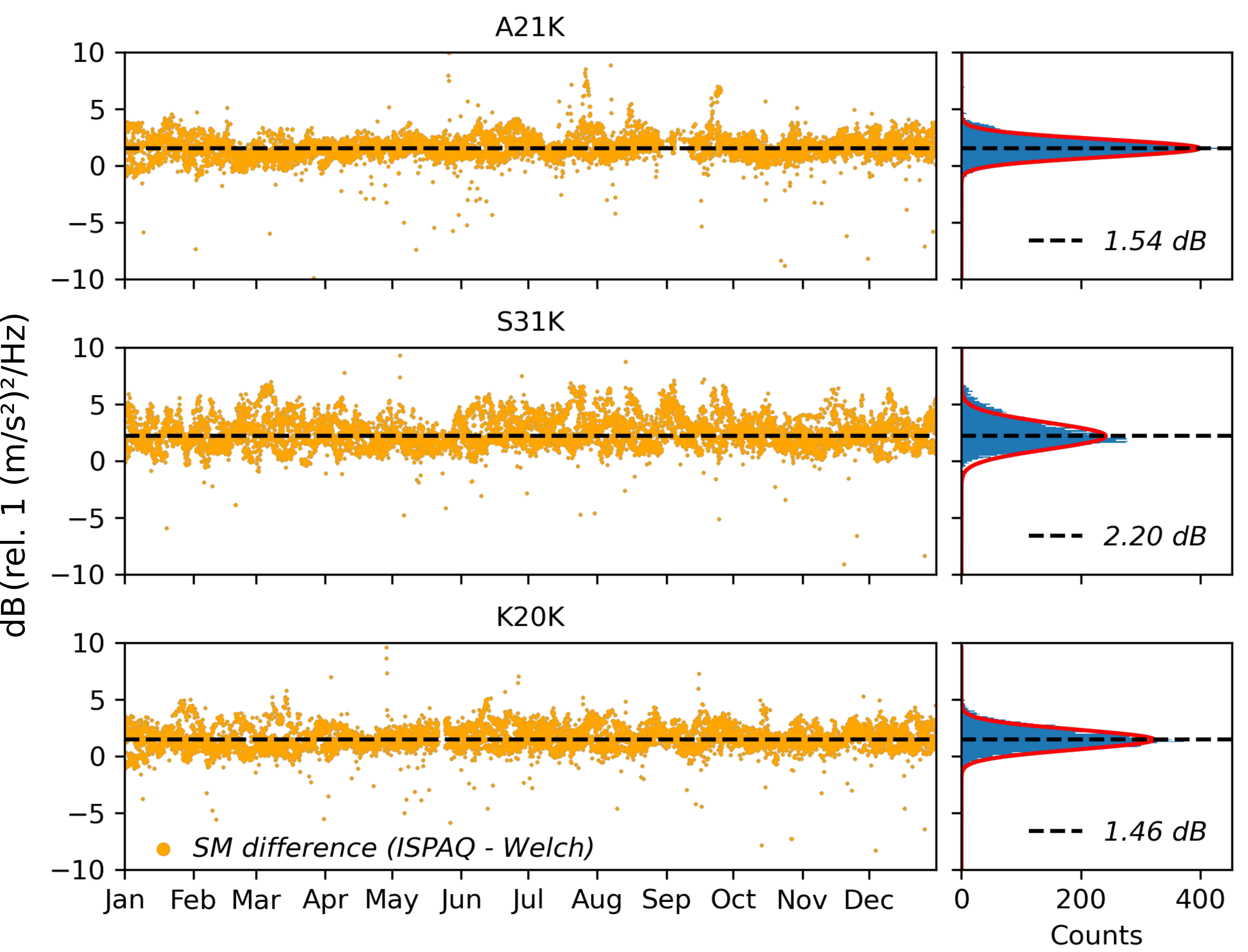}
\caption{\label{fig5}Differences between mean PSDs in the SM (5-10s) band computed using ISPAQ/MUSTANG and Welch's method.}
\end{figure}

\begin{figure}[H]
\centering
\includegraphics[width=1\textwidth]{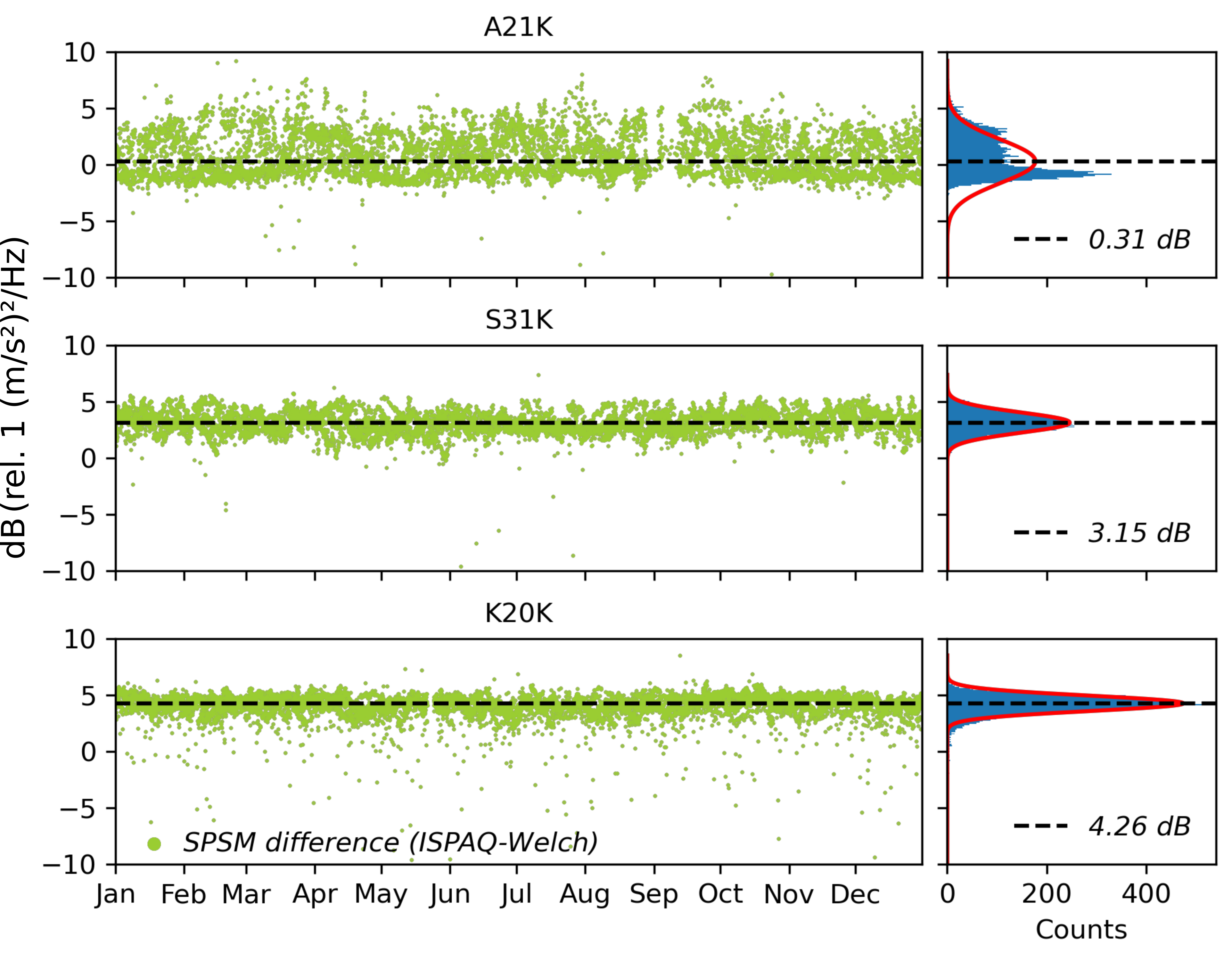}
\caption{\label{fig6}Differences between Mean PSDs in the SPSM (1-2s) band computed using ISPAQ/MUSTANG and Welch's method.}
\end{figure}

\section*{}
\bibliographystyle{johd}
\bibliography{mus}

\end{document}